 \documentclass[preprint]{aastex}
 
\usepackage{graphicx}

\newcommand{\etal}{\mbox{et~al.}}

\def\deg      {{\ifmmode^\circ\else$^\circ$\fi}} 

\slugcomment{Accepted for the Astrophysical Journal, COSMOS Special Issue}
 
 \shorttitle{The COSMOS HST/ACS Observations and Data Processing}
 \shortauthors{Koekemoer et al.}
 
\begin{document}
 
 
 \title{The COSMOS Survey: Hubble Space Telescope / Advanced Camera for Surveys (HST/ACS) Observations and Data Processing\altaffilmark{*}}


 \author{ 
A. M. Koekemoer\altaffilmark{1},
H. Aussel\altaffilmark{2,3},
D. Calzetti\altaffilmark{1},
P. Capak\altaffilmark{4},
M. Giavalisco\altaffilmark{1},
J.-P. Kneib\altaffilmark{5},
A. Leauthaud\altaffilmark{5},
O. Le Fevre\altaffilmark{5},
H. J. McCracken\altaffilmark{6,7},
R. Massey\altaffilmark{4},
B. Mobasher\altaffilmark{1},
J. Rhodes\altaffilmark{4,8},
N. Scoville\altaffilmark{4,9},
P. L. Shopbell\altaffilmark{4}
}

 
\altaffiltext{0}{Based on observations with the NASA/ESA {\em
Hubble Space Telescope}, obtained at the Space Telescope Science
Institute, which is operated by AURA Inc, under NASA contract NAS
5-26555.}
%
%
\altaffiltext{1}{Space Telescope Science Institute, 3700 San Martin
Drive, Baltimore, MD 21218}
\altaffiltext{2}{Institute for Astronomy, 2680 Woodlawn Dr., University of Hawaii, Honolulu, Hawaii, 96822}
\altaffiltext{3}{Service d'Astrophysique, CEA/Saclay, 91191 Gif-sur-Yvette, France}
\altaffiltext{4}{California Institute of Technology, MC 105-24, 1200 East
California Boulevard, Pasadena, CA 91125}
\altaffiltext{5}{Laboratoire d'Astrophysique de Marseille, BP 8, Traverse
du Siphon, 13376 Marseille Cedex 12, France}
\altaffiltext{6}{Institut d'Astrophysique de Paris, UMR7095 CNRS, Universit\`e Pierre et Marie Curie, 98 bis Boulevard Arago, 75014 Paris, France}
\altaffiltext{7}{Observatoire de Paris, LERMA, 61 Avenue de l'Observatoire, 75014 Paris, France}
\altaffiltext{8}{Jet Propulsion Laboratory, Pasadena, CA 91109}
\altaffiltext{9}{Visiting Astronomer, Univ. Hawaii, 2680 Woodlawn Dr., Honolulu, HI, 96822}
%
%

  
\begin{abstract}
We describe the details of the {\it Hubble Space Telescope (HST) Advanced Camera
for Surveys / Wide Field Channel (ACS/WFC)} observations of the COSMOS field,
including the data calibration and processing procedures. We obtained a total
of 583 orbits of HST ACS/WFC imaging in the F814W filter, covering a field
that is 1.64~degrees$^2$ in area, the largest contiguous field ever imaged
with HST. The median exposure depth across the field is 2028~seconds (one HST
orbit), achieving a limiting point-source depth AB(F814W) = 27.2 (5 sigma).
We also present details about the astrometric image registration, distortion
removal and image combination using MultiDrizzle, as well as motivating the
choice of our final pixel scale (30~milliarcseconds per pixel), based on the
requirements for weak lensing science. The final set of images are publicly
available through the archive sites at IPAC and STScI, along with further
documentation on how they were produced.
\end{abstract}
 
 
 \keywords{cosmology: observations --- cosmology: large scale structure of universe --- cosmology: dark matter --- galaxies: formation --- galaxies: evolution --- surveys }
 

 
 \section{Introduction}

The Cosmic Evolution Survey
        (COSMOS: \citealt{scoville2007a})
is the largest contiguous survey ever undertaken with the Hubble Space
Telescope (HST), imaging a $\sim 2$~degree$^2$ equatorial field  with the
primary goal of addressing the coupled evolution of large-scale structure,
star formation and galaxy activity.
A key requirement for this type of cosmological survey is sensitive imaging
at the highest possible angular resolution, particularly for all sources
above $z \sim 0.5 - 1$ where $\lesssim 0\farcs1$ angular resolution becomes
crucial to resolving structure on sub-kpc scales. This approach was pioneered
in the Hubble Deep Fields North and South
	(HDF: \citealt{williams1996,williams2000})
which achieved this level of angular resolution, albeit over a relatively small
5~arcminute$^2$ field. Subsequent surveys included the Hubble Ultra Deep Field
        (UDF: \citealt{beckwith2006})
which went deeper over a somewhat larger 11~arcminute$^2$ area; the Great
Observatories Origins Deep Survey
        (GOODS: \citealt{giavalisco2004})
which was shallower but covered a larger 360~arcminute$^2$ area; and two
larger and shallower surveys, each covering
$\sim 700$~arcmin$^2$
        (GEMS: \citealt{rix2004};
        AEGIS: \citealt{davis2006}).
COSMOS provides an additional factor $\sim10$ increased coverage in area, and
uses broader filters to reach deeper sensitivities which are crucial for
obtaining the accurate galaxy morphology measurements required to construct
mass maps based on weak lensing.

In addition to the HST data, the COSMOS survey also contains a wealth of
ancillary data obtained at other ground-based and space-based telescopes,
including X-ray, infrared, sub-mm and radio imaging, as well as comprehensive
spectroscopic programs, which are all summarized in
	\cite{scoville2007a}.
The present paper describes the details of the HST observations and data
processing that we used to construct the imaging datasets which form the
foundation for the majority of the work in the COSMOS survey; see also
	\cite{scoville2007b}
for an overview.

\section{ACS Observations}

\subsection{ACS Tiling Strategy for the Full COSMOS Mosaic}

The COSMOS survey was awarded a total of 590~orbits of HST observations to
cover a large, contiguous field of ``tiles'' centered at 10:00:28.6, +02:12:21.0
(J2000) using the Advanced Camera for Surveys (ACS) Wide-Field Channel (WFC)
detector, together with the F814W (``Broad I'') filter. The ACS/WFC is a mosaic
detector consisting of two 2048x4096-pixel CCDs with 15$\,\mu$m~pixels,
corresponding to a spatial scale of $\sim 0\farcs05$/pixel and a field of view
of $\sim 202\arcsec \times 202\arcsec$. The detectors have a read noise of
5$\,$e$^-$/pixel and a dark current rate of 0.0038$\,$e$^-$/s/pixel;
the observations were all obtained with a gain of 1, corresponding to a
full-well depth of 84,700 electrons (thus above the maximum A-to-D conversion
limit of 65,535 counts per pixel). The camera is mounted
$\sim 6\arcmin$~off-axis from the principal optical axis of HST and is also
tilted with respect to the focal plane of HST, resulting in a significant
amount of skew distortion across the field, $\sim 7 - 10\%$~of the detector
size. Hence the pixels are projected as skew trapezoids on the sky, with their
shapes changing across the detector, and this distortion is removed during
post-observation processing described further down.

Due to HST scheduling constraints, the observing program had to be divided
between two HST observing cycles from May 2003 to June 2005, with 270~orbits
allocated during Cycle~12 (HST Program ID 9822) and 320~orbits allocated
during Cycle 13 (HST Program ID 10092). In order to maximize the amount of
contiguous area coverage, it was required that all the tiles be obtained at
the same orientation, thereby minimizing the amount of overlap between tiles
to less than a few percent. Combining this with additional constraints on the
orientation of HST with respect to the sun led to all the tiles being tilted
about 10 degrees from North. In order to maintain the edges of the full field
oriented approximately north-south and east-west (to maximize overlap with
ancillary observations from other observatories) this led to slightly
staggered edges, as shown in
        Figure~1.
There are typically up to 16 orbits available per day on HST; since the
allowable roll angles of HST change throughout the year, this meant that not
all the tiles could be obtained at the same orientation, but that some of them
had to be obtained at orientations that were rotated at 180$\deg$
relative to the default orientation. In addition,
approval was granted to use 9 of the 590 orbits to obtain F475W observations
(SDSS $g$) of a $3\times3$~grid at the field center, as a pilot program to
provide a demonstration of the value that two-filter HST ACS/WFC imaging would
have for the main scientific goals of the project. Finally, two of the tile
pointings failed due to problems with guide stars and were repeated, while two
more tiles had severe problems with scattered light from adjacent stars and had
to be repeated using slightly different pointings, offset from the original
pointing by half a field in both directions (for which we used other orbits
from the original total allocation). Thus the final contiguous area is
1.64~degrees$^2$, or approximately 77$\arcmin$~along each side, and contains
data from a total of 583 orbits of HST ACS/WFC F814W imaging (including
2~additional orbits for repeat observations), with an additional 9 orbits
of F475W imaging at the field center.

\subsection{Cosmic Ray Splitting and Dither Strategy for each Tile}

The total ACS/WFC exposure time obtained in F814W for each tile was
2028~seconds. Since cosmic rays impact between $\sim 2 - 6\%$ of the pixels on
the ACS detectors during this length of exposure time, the observations for
each tile were split into four equal-length exposures, each
507~seconds in duration, ensuring that $<1$~pixel out of 4096$^2$ would be
impacted by cosmic rays in all four exposures according to statistical binomial
probability. The exposure time of 507~seconds was the maximum that could be
achieved within the nominal orbital duration of HST, after accounting for
overheads due to readout, telescope motion and guidestar acquisition.

The four 507~second exposures for each tile were obtained using a ``dither''
pattern designed to improve the sampling of the point spread function (PSF),
as well as covering the $\sim$3$\arcsec$~gap that is present between the two CCDs
of the ACS/WFC. In the F814W filter, the intrinsic width of the PSF that is
produced by the HST optics is $\sim 0\farcs085$. However, this is significantly
undersampled by the $0\farcs05$~pixels of the ACS/WFC detector, and the final
measured PSF in the combined images tends to be closer to $\sim 0\farcs1$ as a
result of convolution by the detector pixels as well as the pixel size of the
final image. The undersampling can be mitigated to some extent by offsetting the
detector in a dither pattern that provides subsampling of the PSF in
different parts of each pixel during each exposure. As a consequence of the
changing pixel size produced by the strong distortion of the ACS/WFC detectors,
a given dither offset in arcseconds actually corresponds to different offsets
in terms of pixels, and these offsets gradually vary across the detector,
thereby further modulating the subsampling pattern achieved with four exposures.

We chose a dither pattern designed to produce optimal subsampling of the PSF by
ensuring that any given point would always be sampled by a different part of
each detector pixel in the four exposures. This is achieved by offsetting the
telescope in integer-pixel and half-pixel increments along both the x and y
axes of the detector, thus the combination of four such offsets ensures that
any given point on the sky is sampled by all four quadrants of a pixel during
the four exposures. In addition, we added a $\sim$3$\arcsec$ offset along the
y-direction of the dither pattern in order to cover the gap between the chips,
as well as an offset of a few pixels along the x-direction of the dither
pattern in order to ensure that bad columns and other defects were moved to a
different part of the sky during each of the four exposures. The resulting
dither pattern is shown in
	Figure~2,
demonstrating how we implemented this using a combination of a primary 2-point
dither pattern with an offset of $\sim$6$\arcsec$, together with a secondary
2-point dither pattern at each of the primary dither pattern points, providing
an offset of $\sim$3$\arcsec$ along a slightly different orientation. The dither
offset spacings and orientations in the two patterns provided the required
pattern of half-pixel steps that was needed to ensure good subsampling to the
level of $\sim$0.5~pixel across the entire detector. In
	Figure~3
we show the effective exposure time obtained after combining all four exposures
for an example tile pointing, demonstrating the relatively uniform coverage
across the detectors. Specifically, bad columns are moved around along the
x-axis of the detectors, thereby ensuring that these points on the sky are
covered by good pixels in three other exposures, and in addition the gap
between the chips is always successfully covered by at least three exposures.

\subsection{ACS Filter Selection}

The primary scientific goal of the COSMOS HST ACS/WFC observations is to
obtain the best possible rest-frame optical morphological information on
$z \gtrsim 1$ galaxies, thereby necessitating observations at red wavelengths.
We chose the broadest available filter on ACS in this wavelength range, namely
the F814W (``Broad I'') filter. This filter is characterized by having a
combination of exceptionally high transmission ($>90 - 95\%$) across an
extremely wide wavelength range ($\sim\,$7300\AA$\,-\,$9500\AA); coupled with
the red-sensitive ACS/WFC CCDs this provided optimal sensitivity to faint
rest-frame visible morphological information at the redshifts of interest
($z \gtrsim 1$).

More specifically, the use of the F814W filter provides an additional
$\sim 0.5\,$magnitude of depth relative to the narrower F775W (SDSS $i$)
filter, for a given exposure time. This is important when comparing the
depth of COSMOS to other surveys such as GEMS
        \citep{rix2004},
GOODS
        \citep{giavalisco2004}
or the UDF
        \citep{beckwith2006},
all of which used the SDSS filter set, in particular F775W (SDSS $i$) and
F850LP (SDSS $z$) at the red end of the spectrum, to provide sharp color
discrimination through the use of certain spectral break features. This comes
at the cost of a shallower depth per unit exposure time compared with the F814W
filter, even though these other fields might be observable for a larger fraction
of time with HST, since the total system throughput with the F814W filter is
approximately equal to the sum of the F775W and F850LP filters. For this
reason we chose the F814W filter for the COSMOS survey, to ensure the deepest
possible morphological coverage with HST, while photometry using the SDSS
filter set was obtained from ancillary ground-based observations using Subaru
        \citep{taniguchi2007}.

For the pilot observations of the central $3\times3$ tiles using a bluer filter,
we chose the F475W (SDSS $g$) filter, motivated by the need to separate
extinction-related effects from shape changes introduced by weak lensing. The
F475W filter provides optimal throughput in the rest-frame near-UV for our
target galaxies at $z \gtrsim 1$, for which weak lensing studies are the
primary goal of the survey.

\section{Data Reduction and Processing}

We processed all the ACS data at STScI on a special-purpose computing cluster
purchased especially for the project, consisting of 6 linux CPU nodes running
pipeline scripts in parallel. As each observation was obtained, the individual
exposures were delivered to the computing cluster where they were run through
an IRAF/STSDAS pipeline that performed calibration, astrometric registration
and cosmic ray cleaning, as well as final mosaic combination using
{\tt MultiDrizzle}
        \citep{koekemoer2002},
which makes use of the {\tt Drizzle} software
        \citep{fruchter2002}
to remove the geometric distortion and map the input exposures onto a rectified
output frame. Here we describe the details of each of these steps.

\subsection{Initial Data Calibration}

Each of the individual 507~second exposures was first run through the basic
steps of ACS calibration using the IRAF/STSDAS task {\tt calacs}, which
performs bias and dark subtraction, gain correction, flat fielding, and
identification of bad pixels. Each dataset was retrieved and run through this
pipeline as soon as possible after the observation in order to make the data
available quickly; this generally necessitated a second-pass calibration a
few weeks later with more accurate reference files, in particular using dark
and bias reference files that could be created using dark and bias exposures
obtained contemporaneously with the data.

After basic calibration, several additional effects also had to be
corrected in the data. In many cases, low-level residual background was
present in the images, typically due to scattered light, and this was removed
by constructing a master scattered-light image for the entire dataset through
medianing, and then scaling and subtracting this from each individual exposure.
In addition, the ACS/WFC CCDs are read out by four amplifiers (two for each
CCD), and their bias levels often vary by a few tenths of a count which is not
accounted for in the overscan bias calibration, therefore this was measured
and corrected for each exposure.

Finally, the charge transfer efficiency (CTE) of the detectors is gradually
deteriorating over time, as a result of charge traps in the pixels created by
cosmic ray hits. This results in lost flux as the charge from each pixel is
transferred down the CCD columns during readout, as well as producing trails
as the traps release their charge after it has passed through. This effect is
most severe for pixels that are furthest from the amplifiers, and for faint
sources with minimal sky background. Its severity is reduced for sources on
higher backgrounds (since there are a finite number of traps in a given pixel,
thereby losing a progressively smaller fraction of the flux for brighter
pixels). Ideally this effect would be corrected for on a pixel-by-pixel basis
in the raw images but this requires a complete physical understanding of the
effect, which is still under development. However, the effect can be quantified
sufficiently well that it can be used to apply a post-processing correction to
measured morphological and photometric properties of sources in the images,
particularly relevant to the COSMOS weak lensing studies
        \citep{rhodes2007},
based on a knowledge of the source positions and fluxes, together with the sky
background values for the exposures involved.

\subsection{Astrometric Image Registration}

After calibration, the next step involved registering the images onto an
astrometric grid. The process of aligning the astrometry of HST ACS/WFC images
can be separated into two components: (1)~ensuring that all four exposures for
each tile, obtained during a single orbit, are aligned, and (2)~ensuring that
adjacent tiles are aligned. Given the small angular scale of the ACS/WFC
pixels ($0\farcs05$), it is crucial to align images to better than
$\sim 2 - 5$~milliarcseconds in order to achieve accurate cosmic ray rejection
among the separate exposures within a tile, and to ensure accurate combination
of pixels in overlapping regions between adjacent tiles. In addition, it is
important to accurately remove the strong distortion that is present in the
ACS/WFC images ($\sim 7 - 10\%$); this was done using the MultiDrizzle
software
        \citep{koekemoer2002}
using distortion solutions provided by
        \cite{anderson2005},
which are accurate to better than $\sim 0.05 - 0.1$~pixel across the 4096$^2$
pixel extent of the ACS/WFC.

Since each tile was observed during a single orbit, with dither offsets less
than $\sim 9\arcsec$ in total, there was no need for the telescope to change
guidestars during the orbit. HST generally requires two guidestars for
accurate tracking; if both guidestars were acquired successfully at the start
of the orbit then those are retained throughout the orbit, and the resulting
positional accuracy of small dither offsets is generally known to be better
than $\sim 2 - 3$~milliarcseconds. This was verified by our processing pipeline
which measured the locations of sources on each of the four exposures obtained
during each orbit and compared the resulting calculated shifts with those
which were commanded. It was verified that the r.m.s. accuracy of relative
offsets within an orbit is typically $\lesssim 2$~mas ($\lesssim 0.04$~pixels),
hence there was generally no significant correction required for these relative
offsets within an orbit.

However, a larger astrometric correction is required when aligning adjacent
tiles, which is fundamentally due to limitations in the knowledge of the
positions of the guidestars. During the cycles when the COSMOS data were
obtained (Cycles 12 and 13), all the guidestar positions were still based on
the Guide Star Catalog v1.0 which had no correction for proper motion of
guidestars (improved guidestar positions were incorporated into the system
after Cycle~14), in addition to many other errors being present in the guidestar
catalog. These effects generally lead to an uncertainty of $\sim 1 - 2\arcsec$,
or more, in the guidestar positions, which is a well-known problem for HST data
obtained before Cycle~14. Since the guidestar positions are used to calculate
the astrometric information of the exposures, this means that the absolute
astrometry of each tile could initially be in error by up to
$\sim 1 - 2\arcsec$ or more, even though the relative alignment of the four
exposures of each tile were accurate to $\lesssim 2$~mas. Moreover, since the
two guidestars are typically observed in two different Fine Guidance Sensors (FGSs) separated by $\sim 25\arcmin$, the uncertainty in their absolute
astrometry can translate into an error in the calculated orientation of HST
by as much as $0.01 - 0.03$~degrees in the worst cases, corresponding to a
few pixels across the scale of the ACS/WFC detectors in addition to the error
introduced by the uncertainty in the absolute astrometry.

In order to correct the absolute astrometry of the HST ACS/WFC datasets, we
defined a COSMOS astrometric grid
	\citep{aussel2007}
based on two ancillary datasets: (1)~the first epoch of VLA imaging
        \citep{schinnerer2004},
to provide a robust fundamental astrometric frame by means of a sufficiently
large number of unresolved radio sources across the entire field; (2)~the CFHT
$i^*$~band imaging dataset
        \citep{capak2007},
which was placed onto the VLA astrometric frame by matching unresolved sources
detected in both datasets. The CFHT data in turn provided the best combination
of depth and area to enable large numbers of sources to be identified on each
ACS/WFC tile for astrometric correction.

On each ACS/WFC tile, there were generally a total of $\sim 300 - 600$ sources identified that were also detected on the CFHT $i^*$~band image. The measured
RA and Dec positions of the sources were cross-correlated and matched, in order
to solve for the transformation between them. Since all the higher-order
components of the ACS/WFC distortion are removed by the {\tt MultiDrizzle}
software
        \citep{koekemoer2002},
the ACS/WFC tiles can be treated as rectified images, with the only remaining
unknown terms being purely linear transformations, specifically an offset in
RA,Dec combined with a possible small rotation, due to the uncertainties in the
guidestar positions. Given that the typical astrometric uncertainty in the
CFHT images is $\sim 0.1\arcsec$ for a single source, this can be reduced in
quadrature by using as many source as possible to solve for the transformations,
yielding a combined accuracy of $\sim 5$~milliarcseconds for the resulting
shifts when we used the full set of $\sim 300 - 600$ sources on each ACS image
that had counterparts in the CFHT images.

The above process ensured that all the ACS/WFC exposures and tiles were
registered with an absolute astrometric accuracy of $\sim 5\,$milliarcseconds
across the entire COSMOS field. These images were placed onto an astrometric
frame defined as a tangent plane projection, centered at the nominal COSMOS
pointing of 10:00:28.6, +02:12:21.0 (J2000). The tangent plane projection is
the standard projection used for all astronomical images that have a uniform
projected pixel size on the sky across the entire image; one of its consequences
is that the projected location of a source in the image is slightly further away
from the center than its true angular separation on the sky. For example, at a
distance of $30\arcmin$ from the center, the projected distance of a source in
the image is $0\farcs046$ further away from the center than its true distance.
Since all the ancillary COSMOS datasets also use the tangent plane projection, this effect cancels out and is only relevant when comparing the measured
position of a source in the COSMOS field to other data that are not on the
COSMOS tangent plane, in which case the effect due to the tangent plane
projection can be accounted for using simple trigonometry. Throughout this
paper, we assume use of the tangent plane projection as described here.

\subsection{Cosmic Ray Rejection}

Cosmic ray rejection was carried out primarily among the four exposures
within each tile; adjacent information from overlapping tiles was generally
not used due to the small amount of overlap. Since each exposure was dithered
to a different position on the sky, simple stacking was inadequate to remove
cosmic rays so we used {\tt MultiDrizzle} to perform the cosmic ray rejection.
This process first used {\tt Drizzle} to remove the distortion from each of
the input exposures, mapping them onto four separate output images that were
aligned to the same pixel grid and are rectified, with all the distortion
removed, so that all the pixels subtended the same area on the sky.

Once the four drizzled images were created, we used {\tt MultiDrizzle} to
combine them using a median process to create a clean approximation to the
final output image. For each 507~second exposure, $\sim 80,000 - 250,000$~pixels
out of 4096$^2$ were affected by cosmic rays, therefore it was rare for a pixel
to be affected by cosmic rays during all four exposures ($\ll 1$ pixel out of
4096$^2$). However, the number of pixels affected by only 3~cosmic rays out of
4~exposures was significantly higher (up to $\sim 60$ pixels out of 4096$^2$),
and in addition there were many cases where 2 pixels were affected by cosmic
rays, while the third may lie on a chip defect or the gap between the chips. In
such cases where 3 pixels were affected by a defect while the fourth remaining
pixel was valid, the median was replaced by the value of the valid pixel if the
median value exceeded the valid pixel value by a 5$\sigma$ threshold. This
process successfully ensured that we minimized the number of pixels in the
resulting median image that might be affected by cosmic rays or chip defects.

Once the clean median image was created for each tile, it was then transformed
back to the distorted CCD frame of each of the input exposures, to create a
clean version of the input exposure aligned to the original CCD pixel grid.
Cosmic rays were then identified using sigma-clipping, by comparing the input
exposure with the median and calculating the full r.m.s. by combing the
source counts, background sky counts and read noise in quadrature. In order
to avoid clipping bright stars, the sigma-rejection criterion was softened
using a derivative image, which was constructed from the median such that the
value of each pixel represents the largest gradient from that pixel to its
surrounding neighbors. Pixels are then only rejected as cosmic rays if the
difference between the input exposure and the median exceeds the sum of the
sigma-criterion and the derivative, scaled by an appropriate factor (thus
the derivative component is only significant in bright unresolved sources where
it prevents the cores from being clipped, and becomes insignificant in extended
or faint sources). The rejection was done in two passes, with the first pass
using a 4$\sigma$ clipping combined with a scale factor of 1.2 for the
derivative image. The second pass was only performed on pixels surrounding
cosmic rays identified during the first pass and was aimed at rejecting fainter
pixels associated with the cosmic ray that were not rejected on the first pass,
so a more stringent clipping criterion was used (3$\sigma$, combined with a
scale factor of 0.7). The values for this clipping procedure were arrived at
through extensive exploration of parameter space, and are detailed further in
        \cite{koekemoer2002}.

\section{Construction of the Final Combined Tiles and Mosaic Images}

Once the cosmic ray masks had been created, we then used {\tt MultiDrizzle}
again, this time combining all the input exposures onto a final combined
image. For each exposure, we first created a variance map that contained all
the components of noise in the image except for the sources themselves, thus
containing the sky background (modulated by the flatfield and the geometric
projection of the detector on the sky), the read-out noise, and the dark
current, using the measured values applicable to each of the four amplifiers
on the ACS/WFC chips. The variance image for each exposure was then inverted
and used as a weight map associated with the exposure; inverse variance has
the appealing property of scaling linearly with exposure time, as well as the
ability to exclude bad pixels from the drizzle combination by setting their
inverse variance weight to zero.

The core of the {\tt Drizzle} algorithm consists of transforming each input
detector pixel onto the output image plane, which may have a different pixel
scale and orientation, and distributing the flux from the input pixel among all
the output pixels that it may overlap. The PSF in the resulting image is thus
convolved three times: by the detector pixel scale (in the original exposure),
by the output pixel scale, and again by the detector pixel scale when the
pixels are mapped from the input to the output frame. This third convolution
can be minimized by shrinking the input pixels by an arbitrary factor (the
{\tt pixfrac} parameter) which ranges between 0 and 1; a {\tt pixfrac} value
of 1 corresponds to simple shift-and-add, thus adding a full convolution by
the input pixel size, while a value of 0 corresponds to interlacing where
each input pixel is mapped to a delta function and introduces no additional
convolution in the output image plane. After experimentation with different
values of pixfrac, we chose a value of 0.8 which was well matched to our output
pixel scale and the degree of subsampling achieved by our dither pattern.

We chose an output pixel scale of $0\farcs03$/pixel (0.6 times the input CCD
detector pixel scale) which not only reduced the convolution by the output
pixel size, but also reduced the effects of aliasing that would otherwise
result when the input and output pixel scales are comparable. Reducing the
aliasing effects had the benefit of providing a more stable PSF for the weak
lensing studies; we also used a Gaussian kernel to produce the images for the
lensing studies which further stabilized the PSF, at the expense of adding
some additional correlated noise to the images. For the lensing studies
        \citep{rhodes2007},
we produced a combined version of each tile separately, oriented not with North
up but rather in the default unrotated frame of the ACS/WFC CCDs, to facilitate
processing with the PSF matching procedures.

For the rest of the COSMOS science projects, we ran {\tt MultiDrizzle} again
to produce additional images, this time oriented with North up, using a square
kernel with {\tt pixfrac} set to 0.8, and registered with
$\sim 5$~milliarcsecond precision onto the corresponding images created
from the Subaru and CFHT data
        \citep{capak2007}
to facilitate direct comparison between the HST ACS/WFC data and all the other
ground-based ancillary data for COSMOS. These images were created at two
pixel scales, $0\farcs03$/pixel and $0\farcs05$/pixel; we also used the
latter set of images to create a single monolithic mosaic file extending
for 100,800 pixels along its x and y axes, thus $1.4\deg$ on a side; this
image is displayed in
	Figure~4.
These data products are publicly available through the IPAC/IRSA and
STScI/MAST data archive interfaces.

\section{Summary}

We have presented in this paper the details of the HST ACS/WFC COSMOS
observations and data processing that we used to produce the imaging datasets
which form the basis for the majority of the scientific work in the COSMOS
project. A more general discussion of the HST ACS dataset is presented in
	\cite{scoville2007b}.
The relative astrometry of all the HST images is accurate to
$\sim 5$~milliarcseconds, with an absolute astrometric accuracy determined
fundamentally by the accuracy of the radio reference frame which is
$\sim 55$~milliarcseconds.  The images reach a point-source limiting depth
AB(F814W) = 27.2 (5$\sigma$) in a $0\farcs24$ diameter aperture, and have been
projected onto output grids of $0\farcs03$/pixel and $0\farcs05$/pixel. The
$0\farcs03$/pixel data was drizzled using a Gaussian kernel with a FWHM of
40~milliarcseconds, and the average width of the PSF in these images is
$0\farcs095$; these images are optimized for weak lensing studies, and are
available as unrotated tiles (one for each ACS orbit) and also as combined
sections oriented with North up, registered to the pixel frame of the
ground-based optical datasets. The $0\farcs05$/pixel data was drizzled using a
square kernel with a pixfrac of 0.8 and has an average PSF width of
$0\farcs097$. This dataset is available in sections that are registered to the
pixel frame of the ground-based optical datasets, and is also available as a
single monolithic mosaic 100800 pixels on a side (80 Gb). The COSMOS HST
datasets are publicly available through the web sites for
IPAC/IRSA ({\bf \url{http://irsa.ipac.caltech.edu/data/COSMOS/}})
and STScI-MAST ({\bf \url{http://archive.stsci.edu/}}).
IRSA also supplies a cutout capability derived  from the full field mosaic,
which can be made with any field center and size.

 
 \acknowledgments
 
The HST COSMOS Treasury program was supported through NASA grant
HST-GO-09822. We wish to thank Tony Roman, Denise Taylor, and David 
Soderblom for their assistance in planning and scheduling of the extensive COSMOS observations. We gratefully acknowledge the contributions of the
entire COSMOS collaboration consisting of more than 70 scientists. 
More information on COSMOS is available at
{\bf \url{http://cosmos.astro.caltech.edu/}}
and we acknowledge the services provided by the staff at the NASA IPAC/IRSA
as well as the STScI MAST Archive in providing online archive and server
capabilities for the COSMOS datasets. The COSMOS Science meeting in May 2005
was supported in part by the NSF through grant OISE-0456439.

\clearpage
 
 
 
\begin{figure}[ht]
\epsscale{1.0} 
\plotone{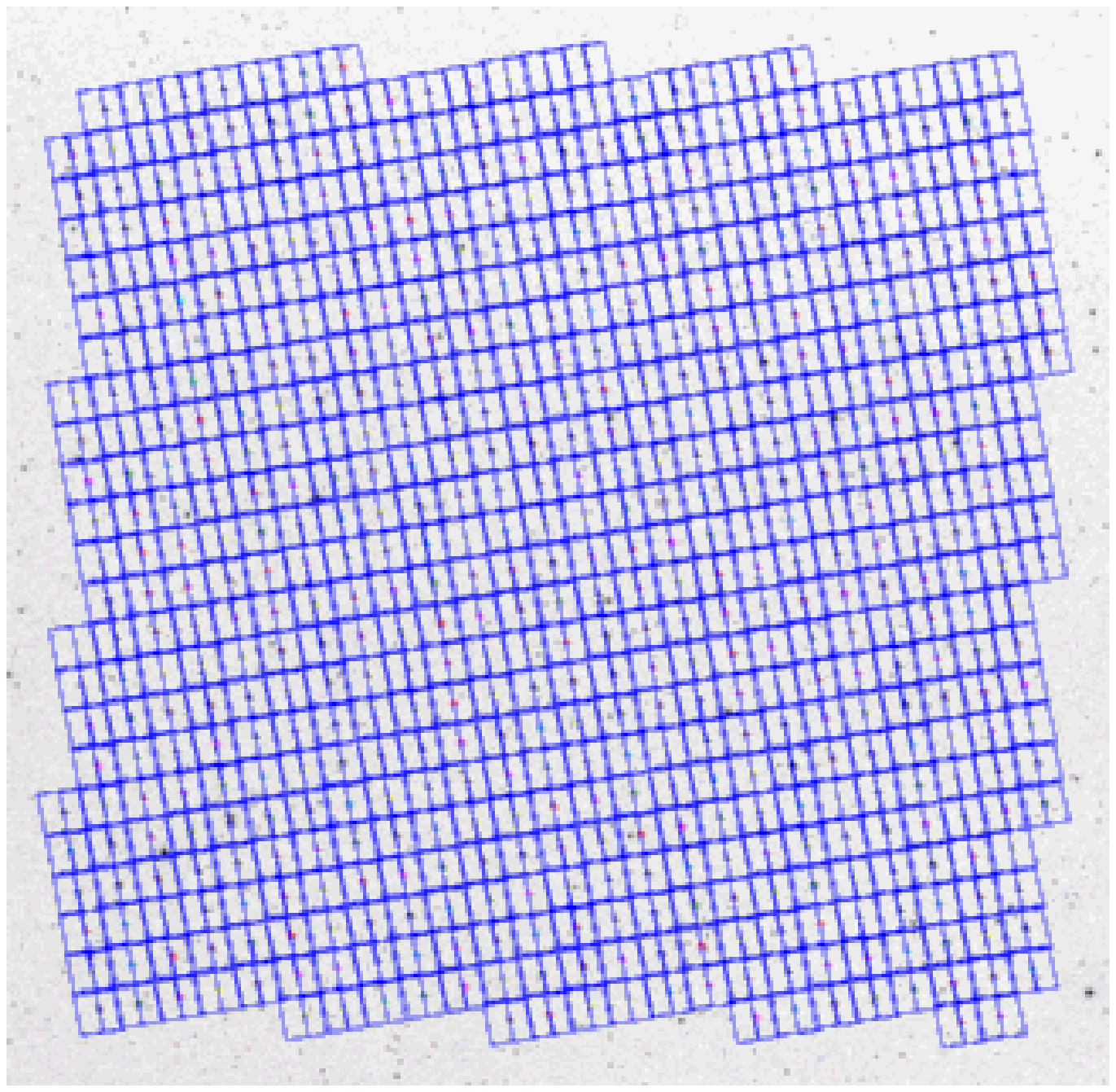}
\caption{The final layout of the full mosaic of all the ACS/WFC F814W pointings,
covering a total area of 1.64~degrees$^2$, or about $77\arcmin$ on a side. The
field is centered at 10:00:28.6, +02:12:21.0 (J2000); the rectangle fully
enclosing all the ACS imaging has lower left and upper right corners
(RA,DEC J2000) at  (150.7988\deg,1.5676\deg) and  (149.4305\deg, 2.8937\deg).} 
\end{figure}

\clearpage
 
\begin{figure}[ht]
\epsscale{1.0} 
\plotone{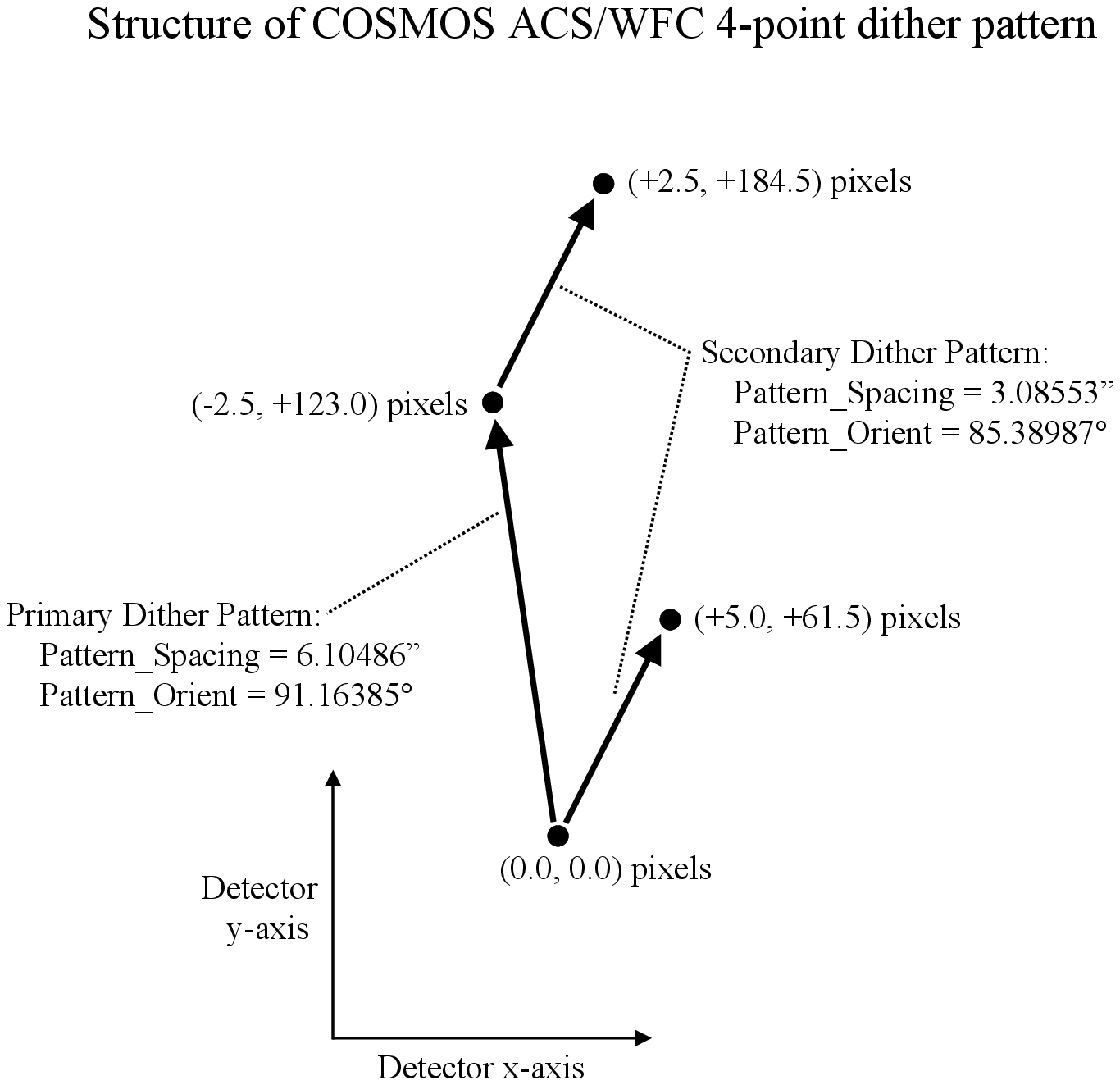}
\vspace{40pt}
\caption{Schematic diagram of the dither pattern that we used for each set of
four exposures that were obtained for each tile. By combining a 2-point
primary dither pattern with a 2-point secondary pattern at each of the two
primary dither points, along a slightly different orientation, each of the
four pointings was thereby separated in increments of $\sim$3$\arcsec$ along the
y-axis of the detector and increments of $\sim0\farcs25$ along the x-axis of the
detector. The {\tt Pattern\_Spacing} and {\tt Pattern\_Orient} parameters shown
are those that were used in the phase~II file to construct these offsets.
This pattern provided a combination of integer-pixel and half-pixel offsets,
thereby ensuring that all four pointings would yield good sampling of any point
on the sky by all four quadrants of a detector pixel. The distorted geometry of
the detector provides an additional modulation of the phase of this half-pixel
subsampling across the image, once all four exposures are combined.} 
\end{figure}

\clearpage
 
\begin{figure}[ht]
\epsscale{1.0} 
\plotone{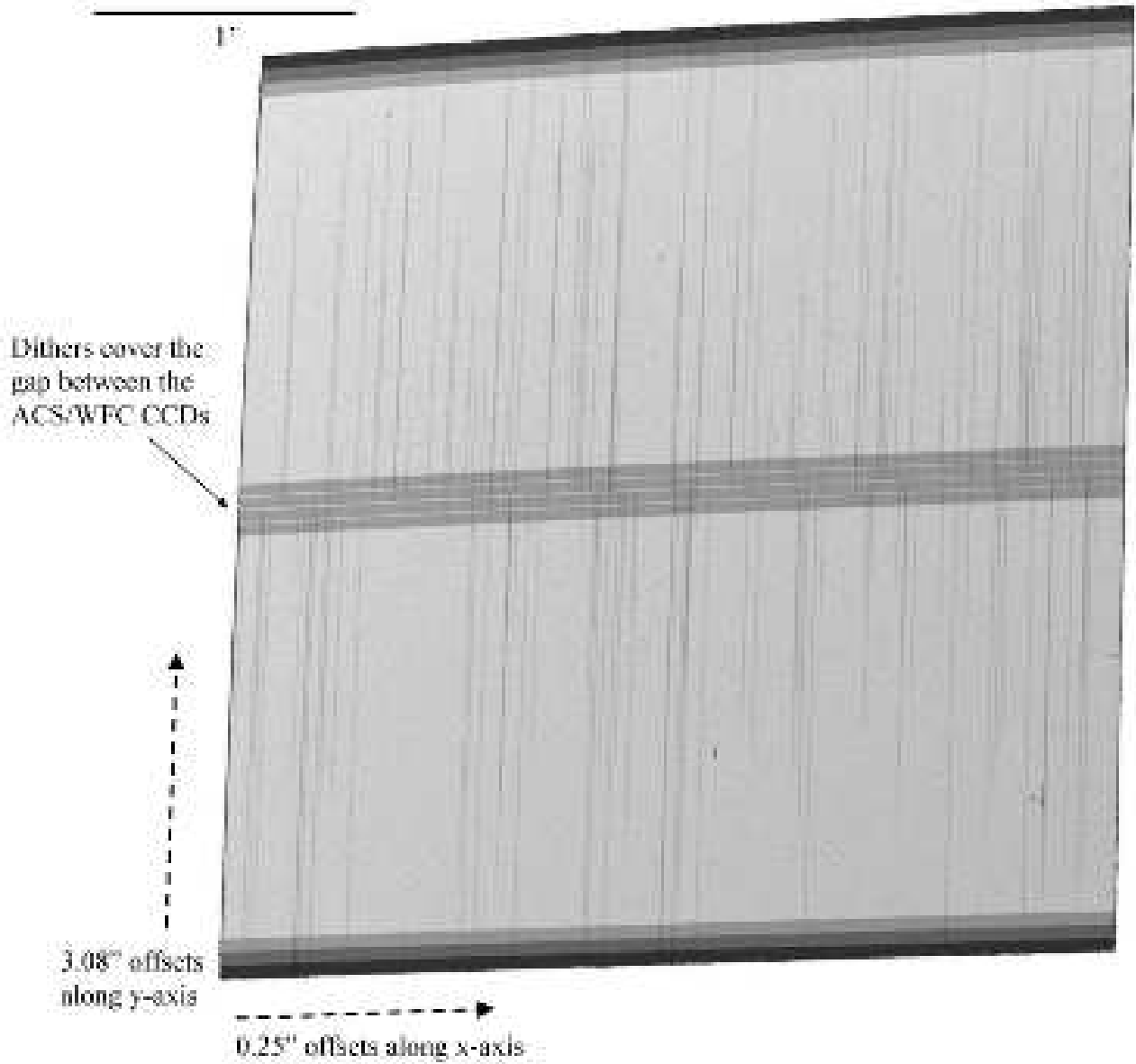}
\vspace{12pt}
\caption{Example of a ``weight'' image corresponding to the combination of
four ACS/WFC exposures of a single pointing, after initial processing through
the pipeline. The image is equivalent to effective exposure time, with lighter
regions indicating a longer combined exposure time on the final output pixels.
Vertical dark lines correspond to bad columns, while the horizontal bands
across the centre of the image correspond to the $\sim$3$\arcsec$ gap between
the two detectors. This image demonstrates that the dither pattern successfully
provided coverage of the gap by at least three exposures, as well as moving
bad columns along the x-direction sufficiently to ensure that they were always
covered by good pixels in the other exposures.} 
\end{figure}

\clearpage
 
\begin{figure}[ht]
\epsscale{1.0} 
\plotone{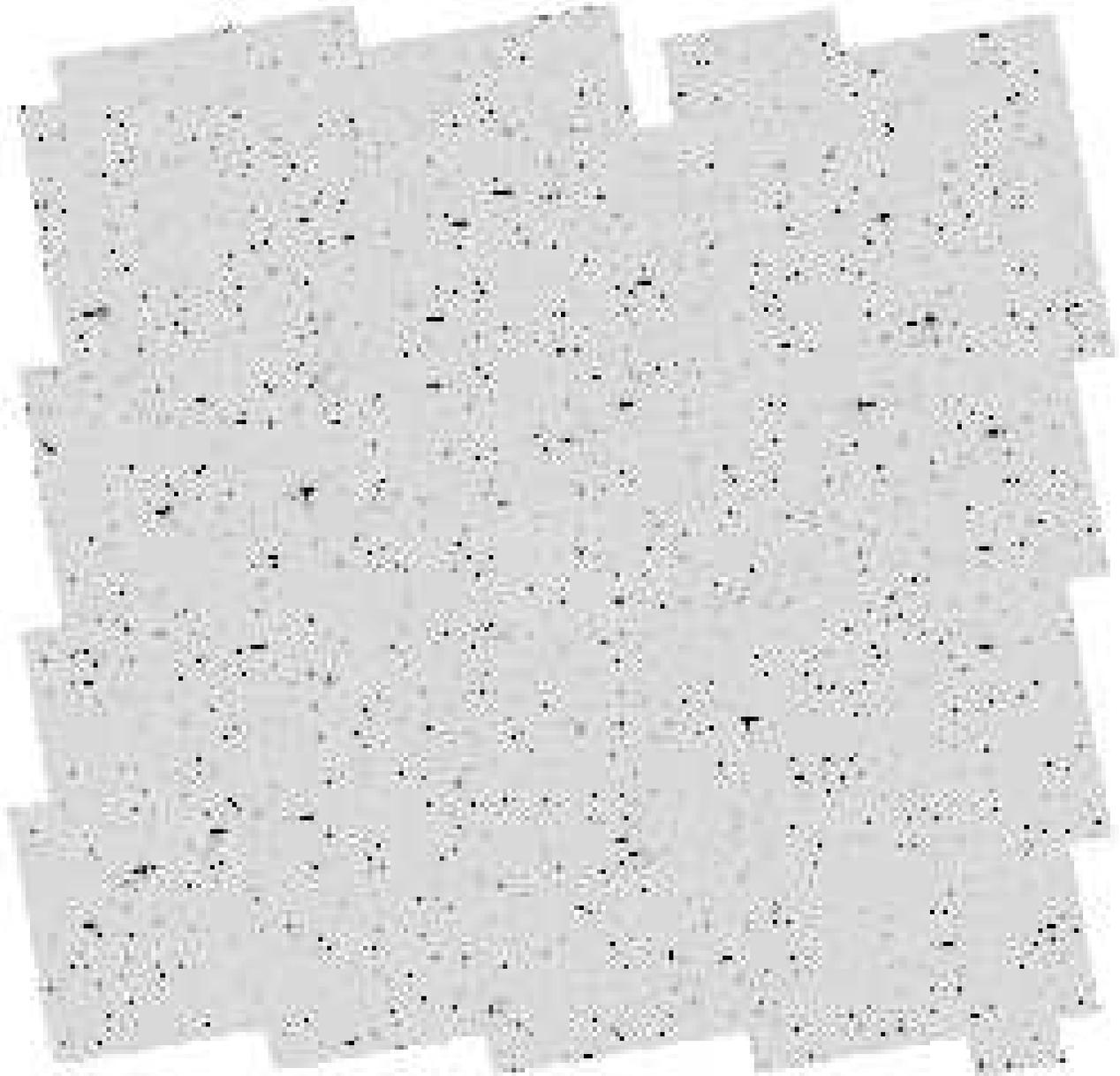}
\caption{The final combined mosaic of all the ACS pointings. At a pixel scale
of $0\farcs05$/pixel, this image has 100800 pixels on a side ($\sim 84\arcmin$),
with the observations covering an area of 1.64~degree$^2$. The field is
centered at 10:00:28.6, +02:12:21.0 (J2000); the rectangle fully enclosing all
the ACS imaging has lower left and upper right corners (RA,DEC J2000) at 
(150.7988\deg,1.5676\deg) and  (149.4305\deg, 2.8937\deg).} 
\end{figure}

\clearpage
 
\begin{deluxetable}{lcc}
\tabletypesize{\scriptsize}
\tablecaption{COSMOS HST ACS/WFC Observations\label{tbl-2}}
\tablewidth{0pt}
\tablehead{
\colhead{Range of dates} & \colhead{PA (Deg)} &  \colhead{Number of pointings}} 
\startdata
2003-Oct-15 --- 2004-Jan-07	& 100	& ~42	\\
2004-Mar-02 --- 2004-May-21	& 280	& 303	\\
2004-Oct-13 --- 2005-Jan-07	& 100	& 103	\\
2005-Mar-02 --- 2005-May-21	& 280	& 142	\\
2005-Oct-28 --- 2005-Nov-24	& 100	& ~~2\tablenotemark{a}
\enddata
\tablenotetext{a}{These two additional orbits were awarded to compensate for
two pointings lost due to guidestar failures.}
\end{deluxetable}

 

\begin{thebibliography}{}
     
\bibitem[Anderson (2005)]{anderson2005}
	Anderson, J. 2005, HST Calibration Workshop, eds. A. M. Koekemoer,
	P. Goudfrooij, L. L. Dressel (Baltimore: STScI), 11

\bibitem[Aussel \etal (2007)]{aussel2007}
	Aussel, H. \etal 2007, \apjs, this volume

\bibitem[Beckwith \etal (2006)]{beckwith2006}
	Beckwith, S. V. W., Stiavelli, M., Koekemoer, A. M., Caldwell, J. A. R.,
	Ferguson, H. C., Hook, R., Lucas, R. A., Bergeron, L. E., Corbin, M.,
	Jogee, S., Panagia, N., Robberto, M., Royle, P., Somerville, R. S.,
	Sosey, M.
	2006, \aj, 132, 1729
     
\bibitem[Capak \etal (2007)]{capak2007}
	Capak, P. \etal 2007, \apjs, this volume

\bibitem[Davis \etal (2006)]{davis2006}
	Davis, M.. \etal. 2006, \apj, submitted


\bibitem[Fruchter \& Hook (2002)]{fruchter2002}
	Fruchter, A. S. \& Hook, R. N. 2002, \pasp, 114, 144

\bibitem[Giavalisco \etal (2004)]{giavalisco2004}
	Giavalisco, M., Ferguson, H. C., Koekemoer, A. M., Dickinson, M.,
	Alexander, D. M., Bauer, F. E., Bergeron, J., Biagetti, C.,
	Brandt, W. N., Casertano, S., Cesarsky, C., Chatzichristou, E.,
	Conselice, C., Cristiani, S., Da Costa, L., Dahlen, T., de Mello, D.,
	Eisenhardt, P., Erben, T., Fall, S. M., Fassnacht, C., Fosbury, R.,
	Fruchter, A., Gardner, J. P., Grogin, N., Hook, R. N.,
	Hornschemeier, A. E., Idzi, R., Jogee, S., Kretchmer, C., Laidler, V.,
	Lee, K. S., Livio, M., Lucas, R., Madau, P., Mobasher, B.,
	Moustakas, L. A., Nonino, M., Padovani, P., Papovich, C., Park, Y.,
	Ravindranath, S., Renzini, A., Richardson, M., Riess, A., Rosati, P.,
	Schirmer, M., Schreier, E., Somerville, R. S., Spinrad, H., Stern, D.,
	Stiavelli, M., Strolger, L., Urry, C. M., Vandame, B., Williams, R.,
	Wolf, C.
	2004, \apjl, 600, 99

\bibitem[Koekemoer \etal (2002)]{koekemoer2002}
	Koekemoer, A. M., Fruchter, A. S., Hook, R. N. \& Hack, W. 2002,
	HST Calibration Workshop, eds. S. Arribas, A. M. Koekemoer, B. Whitmore
	(Baltimore: STScI), 337

\bibitem[Rhodes \etal (2007)]{rhodes2007}
	Rhodes, J. R. \etal 2007, \apjs, this volume

\bibitem[Rix \etal (2004)]{rix2004}
	Rix, H.-W., Barden, M., Beckwith, S. V. W., Bell, E. F., Borch, A.,
	Caldwell, J. A. R., H{\"a}ussler, B., Jahnke, K., Jogee, S.,
	McIntosh, D. H., Meisenheimer, K., Peng, C. Y., Sanchez, S. F.,
	Somerville, R. S., Wisotzki, L., Wolf, C.
	2004, \apjs, 152, 163

\bibitem[Schinnerer \etal (2004)]{schinnerer2004}
	Schinnerer, E., Carilli, C. L., Scoville, N. Z., Bondi, M., Ciliegi, P.,
	Vettolani, P., Le F{\`e}vre, O., Koekemoer, A. M., Bertoldi, F.,
	Impey, C. D.
	2004, \aj, 128, 1974

\bibitem[Scoville \etal (2007a)]{scoville2007a}
	Scoville, N. Z.\etal 2007a, \apjs, this volume

\bibitem[Scoville \etal (2007b)]{scoville2007b}
	Scoville, N. Z. \etal  2007, \apjs, this volume

\bibitem[Taniguchi \etal (2007)]{taniguchi2007}
	Taniguchi, Y. \etal  2007, \apjs, this volume

\bibitem[Williams \etal (1996)]{williams1996}
	Williams, R. E., Blacker, B., Dickinson, M., Dixon, W. V. D.,
	Ferguson, H. C., Fruchter, A. S., Giavalisco, M., Gilliland, R. L.,
	Heyer, I., Katsanis, R., Levay, Z., Lucas, R. A., McElroy, D. B.,
	Petro, L., Postman, M., Adorf, H.-M., Hook, R.
	1996, \aj, 112, 1335

\bibitem[Williams \etal (2000)]{williams2000}
	Williams, R. E., Baum, S., Bergeron, L. E., Bernstein, N.,
	Blacker, B. S., Boyle, B. J., Brown, T. M., Carollo, C. M.,
	Casertano, S., Covarrubias, R., de Mello, D. F., Dickinson, M. E.,
	Espey, B. R., Ferguson, H. C., Fruchter, A. S., Gardner, J. P.,
	Gonnella, A., Hayes, J., Hewett, P. C., Heyer, I., Hook, R., Irwin, M.,
	Jones, D., Kaiser, M. E., Levay, Z., Lubenow, A., Lucas, R. A.,
	Mack, J., MacKenty, J. W., Madau, P., Makidon, R. B., Martin, C. L.,
	Mazzuca, L., Mutchler, M., Norris, R. P., Perriello, B.,
	Phillips, M. M., Postman, M., Royle, P., Sahu, K., Savaglio, S.,
	Sherwin, A., Smith, T. E., Stiavelli, M., Suntzeff, N. B.,
	Teplitz, H. I., van der Marel, R. P., Walker, A. R., Weymann, R. J.,
	Wiggs, M. S., Williger, G. M., Wilson, J., Zacharias, N., Zurek, D. R.
	2000, \aj, 120, 2735

\end{thebibliography}
 \end{document}